\documentclass[aps,nofootinbib,showpacs,twocolumn,floatfix]{revtex4}
\usepackage{graphicx}
\usepackage{epsf,amsfonts,amssymb,amsbsy}
\begin{document}
\title{Entropy of BTZ black hole and its spectrum\\ by quantum radial
geodesics behind horizons}
\author{V.V.Kiselev}
\email{kiselev@th1.ihep.su}
\affiliation{Russian State Research Center ``Institute for High
Energy Physics'', 
Pobeda 1, Protvino, Moscow Region, 142281, Russia\\ Fax:
+7-0967-744937}
\begin{abstract}
In the framework of thermal quantization of radial geodesics
completely confined behind the horizons we calculate the entropy
of BTZ black hole in agreement with the Bekenstein--Hawking
relation. Particles in the BTZ black hole occupy the only quantum
ground level. The quantization allow us to find a linear
dependence of black hole mass versus its orbital momentum.
\end{abstract}
\maketitle


\section{Introduction}
Recently, we have formulated a new framework for calculating the
entropy of black hole by a quasi-classical description of
particles moving at radial geodesics completely confined behind
horizons of black hole \cite{K1,K2,K2',K3}. The thermodynamical
assignment of such the motion is twofold. First, the trajectories
are periodic in purely imaginary time, which is a direct
indication of thermal equilibrium in terms of path integral for
the partition function. The periodicity leads to a thermal
quasi-classical quantization, so that the partition function is
approximated by a leading term of sum over the discrete geodesics.
Second, the macro-state of black hole is fixed by its mass and
other several global charges as an electric charge or an orbital
momentum, while the micro-state is composed of full set of quantum
levels for each particle at confined geodesics. Therefore, the
macro-state is not a pure quantum state, it is characterized by a
density matrix obtained by summing up over the micro-states. Such
the approach has been tested on standard examples of black holes,
i.e. the Schwarzschild black hole, and Reissner--Nordst\"om,
Kerr--Newman ones \cite{K2}. The Hawking radiation
\cite{Hawkingradiation} of Schwarzschild black hole by transitions
between the levels of particle confined behind the horizon has
been investigated in \cite{K3}.

In present paper we expand the analysis to the
Ba\~{n}ados--Teitelboim--Zanelli black hole \cite{BTZ}.

The motion of massive particles in the gravitational field of BTZ
black hole can be studied by the action given by the sum of
gravity and particles terms
\begin{equation}\label{GRaction}
    S=\frac{1}{16\pi G}\int\sqrt{-g}\,{\rm
    d}^3x\,\left\{R+\frac{2}{\ell^2}\right\}-\sum mc\int{\rm d}s,
\end{equation}
where $R$ is the Ricci scalar of curvature, $G$ is the
gravitational constant with the dimension of length or inverse
energy, the length $\ell$ represents the cosmological constant of
anti-de Sitter (AdS) 2+1 space-time, while the interval of BTZ
black hole is given by
\begin{equation}\label{BTZ}
    {\rm d}^2s =N^2{\rm d}^2t-\frac{1}{N^2}\,{\rm
    d}\rho^2-\rho^2\left(-\frac{4J\, G}{\rho^2}\,{\rm d}t+{\rm
    d}\phi\right)^2.
\end{equation}
In (\ref{BTZ})
\begin{equation}\label{N}
    N^2=\frac{1}{\ell^2\rho^2}\,\big(\rho^4-8\,G\,M\ell^2\rho^2+16J^2G^2\ell^2\big),
\end{equation}
whereas $J$ is the orbital momentum of black hole.

The BTZ black hole has two horizons (nullifying the function
$N^2$)
\begin{equation}\label{2hor}
    \rho^2_\pm=4\,G\,M\,\ell^2\left(1\pm\sqrt{1-\frac{J^2}{M^2\ell^2}}\,\right),
\end{equation}
determining two 1D `areas'
\begin{equation}\label{2area}
    {\cal A}_\pm=2\pi\,\rho_\pm.
\end{equation}
The solution is the black hole, if $\rho_\pm$ are real numbers,
i.e.
\begin{equation}\label{limit}
    |J|\leqslant M\,\ell.
\end{equation}

In section II we study the BTZ black hole with $J=0$. We in detail
describe the introduction of appropriate coordinates for geodesics
completely confined behind the horizon and the thermodynamical
quantization, which results in the only ground level of particle
in the field of BTZ black hole. Fixing the total energy of
particles by the black hole mass, we derive the entropy satisfying
the Bekenstean--Hawking law. The case of $J\neq 0$ with two
horizons is investigated in section III. Consistence of two maps
behind the horizons leads to quantizing the ratio of horizon
radii. The ratio has to run a finite set of values, since we
introduce a winding number equal to a number of cycles per period
in imaginary time, and it should be integer. The entropy of BTZ
black hole in general case is reproduced. Section IV represents
the quasi-classical spectrum of BTZ black hole: the mass is linear
in $J$, and the area squared is also linear in $J$. Our
conclusions are listed in section V.

\section{No rotation: $\boldsymbol{J=0}$}
At $J=0$ the metric is reduces to
\begin{equation}\label{BTZ0}
    {\rm d}s^2=N_0^2{\rm d}t^2-\frac{1}{N_0^2}\,{\rm d}r^2-r^2\,{\rm
    d}\phi^2,
\end{equation}
with
\begin{equation}\label{N0}
    N_0^2=\frac{r^2-r_+^2}{\ell^2},\qquad r_+^2=8\,G\,M\,\ell^2.
\end{equation}
At radial geodesics
\begin{equation}\label{rdot}
    (\dot r)^2=N_0^4(1-A\,N_0^2),
\end{equation}
where the integral of motion $A$ is determined by the total energy
of particle, $E$, with the mass $m$,
\begin{equation}\label{A}
    A=\frac{m^2}{E^2}.
\end{equation}
The interval at the trajectory is expressed by
\begin{equation}\label{ds0}
    {\rm d}s^2=\frac{A}{1-A\,N_0^2}\,{\rm
    d}r^2=\frac{\ell^2}{r_c^2-r^2}\,{\rm d}r^2,
\end{equation}
where
\begin{equation}\label{rc}
    r_c^2=r_+^2+\frac{\ell^2}{A}.
\end{equation}
For the radial geodesics completely confined behind the horizon,
we get
\begin{equation}\label{conf}
    r_c^2<r_+^2\quad\Rightarrow\quad A<0,
\end{equation}
so that
\begin{equation}\label{cond}
    {\rm d}s^2>0,\qquad\mbox{if $r$:}\quad 0<r<r_c<r_+.
\end{equation}
At such causal geodesics
\begin{equation}\label{dst}
    {\rm d}s^2=A\,N_0^4\,{\rm d}t^2>0,\quad \mbox{at }
    A<0,\quad\Rightarrow\quad {\rm d}t^2<0,
\end{equation}
i.e. the time should run purely imaginary values. Such evolution
in the imaginary time can be assigned to the thermodynamical
character of motion. Namely, a period in imaginary time is equal
to inverse temperature of thermodynamical ensemble in a thermal
equilibrium. In order to fix the period let us to consider the
Kruskal coordinates.

First, the `tortoise' coordinate is determined by
\begin{equation}\label{tort}
    r_*=\int\frac{{\rm
    d}r}{N_0^2}=\frac{\ell^2}{2r_+}\ln\frac{r-r_+}{r+r_+},
\end{equation}
while the Kruskal coordinates are
\begin{equation}\label{i1}
\left\{\begin{array}{l} \vspace*{2mm}
 u=t-r_*,\\
 v=t+r_*,\end{array}
 \right.\qquad\Rightarrow\qquad
 \left\{\begin{array}{l}
\vspace*{-5mm}
\\
 \displaystyle\bar u=-\frac{\ell^2}{r_+}\,
e^{\displaystyle-u\,\frac{r_+}{\ell^2}}
,\\[4mm]
\displaystyle\bar v=+\frac{\ell^2}{r_+}\,
e^{\displaystyle+v\,\frac{r_+}{\ell^2}},
\end{array}
 \right.
\end{equation}
that results in the metric
\begin{equation}\label{Kru0}
    {\rm d}s^2=\frac{(r+r_+)^2}{\ell^2}\,{\rm d}\bar u\,{\rm
    d}\bar v - r^2\,{\rm d}\phi^2,
\end{equation}
which is free of any singularity, in particular, at the horizon.
The inverse transition is given by
\begin{equation}\label{inverseKru}
    \left\{\begin{array}{l}\vspace*{-5mm}\\
\displaystyle t= \frac{\ell^2}{2r_+}\ln\left[-\frac{\bar v}{\bar u}\right],\\[5mm]
\displaystyle r_* =\frac{\ell^2}{2r_+}\ln\left[-{\bar u\,\bar
v}\cdot\frac{r_+^2}{\ell^4}\right].
\end{array}
 \right.
\end{equation}
Metric (\ref{Kru0}) is useful for analyzing the causal relations,
especially, in the radial motion, since isotropic lines, i.e.
light cones (${\rm d}s^2=0$), correspond to lines parallel to axes
of $\bar v$ and $\bar u$ (${\rm d}\bar v=0$ or ${\rm d}\bar u=0$).

Second, we introduce the coordinate map for the geodesics
completely confined behind the horizon in terms of $\rho$ and
phase $\varphi_E$
\begin{equation}\label{eucliddef}
    \left\{\begin{array}{l}
 \bar u=\,\varkappa\,{\rm i}\,\rho\,e^{{\rm i}\,\varphi_E},\\[2mm]
 \displaystyle
 \bar v=-\frac{{\rm i}}{\varkappa}\,\rho\,e^{-{\rm i}\,\varphi_E},\end{array}
 \right.
\end{equation}
which are related with the primary time and radius by
\begin{equation}\label{euclid-def2}
    \left\{\begin{array}{rlrl}
 t=& \displaystyle -{\rm i}\,\frac{\ell^2}{r_+}\,\varphi_E+\Delta t_0,
   & \varphi_E\in & [0,2\pi],
 \\[6mm]\displaystyle
 r_*=& \displaystyle \frac{\ell^2}{r_+}\, \ln\left[-\frac{\rho\,r_+}{\ell^2}\right],
     &\rho \in &[0,\ell^2/r_+],\end{array}
 \right.
\end{equation}
 at $\Delta t_0=-\ell^2/r_+\,\ln\varkappa$.
Eq. (\ref{euclid-def2}) certainly establishes the value of period
in the purely imaginary time, that determines the inverse
temperature of black hole at $J=0$,
\begin{equation}\label{invT0}
    \beta_0=2\pi\,\frac{\ell^2}{r_+}.
\end{equation}

The horizon `area' (perimeter) is given by
\begin{equation}\label{Area0}
    {\cal A}_0=2\pi\,r_+=4\pi\,\ell\sqrt{2GM},
\end{equation}
where we have substituted $r_+$ in terms of black hole mass at
$J=0$. Then, by differentiation we find
\begin{equation}\label{dMdA}
    {\rm d}M=\frac{T_0}{4G}\,{\rm d}{\cal A}_0,\qquad
    T_0=\frac{1}{\beta_0},
\end{equation}
that indicates the introduction of entropy by
\begin{equation}\label{S0}
    {\cal S}_0=\frac{{\cal A}_0}{4G},
\end{equation}
in accordance with the Bekenstein--Hawking relation
\cite{Bekenstein,Hawkingradiation,Hawkingentropy}.

\begin{figure}[th]
\centerline{  \includegraphics[width=8cm]{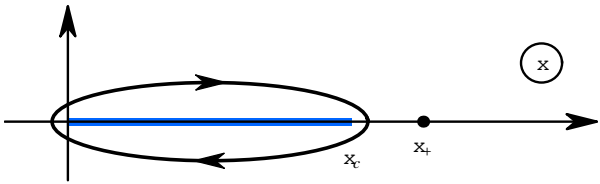}}
  \caption{The integration contour in complex plane of $\xi$.}\label{pic-loop1}
\end{figure}

In order to calculate the partition function of particles moving
by classical paths completely confined behind the horizon
\cite{K1}, we, first, evaluate the increment of phase per cycle,
i.e. by motion from $r=0$ to $r=r_c$ and back,
\begin{equation}\label{inc-p0}
\begin{array}{rl}
    \Delta_c\varphi_E= &
    \displaystyle {\rm i}\cdot 2\int\limits_0^{r_c}{\rm
    d}r\,\frac{{\rm d}t}{{\rm
    d}r}\,\frac{r_+}{\ell^2}=\\[5mm] &
    \displaystyle 2\int\limits_0^{r_c}\,{\rm
    d}r\,\frac{r_+}{r_+^2-r^2}\,\sqrt{\frac{r_+^2-r_c^2}{r_c^2-r^2}}.
    \end{array}
\end{equation}
Introducing
$$
    \xi=r^2,
$$
we can represent integral (\ref{inc-p0}) by
\begin{equation}\label{inc-p01}
    \Delta_c\varphi_E=r_+ \oint \frac{{\rm
    d}\xi}{2\sqrt{\xi}}\,\frac{1}{\xi_c-\xi}\,\sqrt{\frac{\xi_+-\xi_c}{\xi_c-\xi}},
\end{equation}
where the contour of integration loops the cut in complex plane of
$\xi$ because of square roots in the integrand as shown in Fig.
\ref{pic-loop1}.

\begin{figure}[th]
\centerline{  \includegraphics[width=8cm]{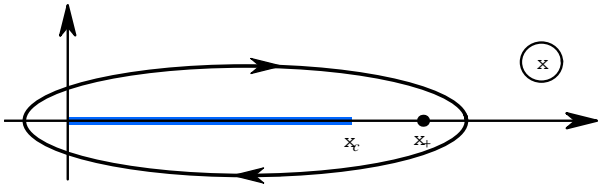}}
  \caption{The auxiliary contour in complex plane of $\xi$.}\label{pic-loop2}
\end{figure}

However, we can consider the auxiliary contour including the pole
at $\xi_+$ in Fig. \ref{pic-loop2}.

\noindent
 This contour can be deformed to the circle at infinity, where the
 integrand rather rapidly nullifies. Therefore, the integral over the auxiliary
 contour is equal to zero, which can be written as the sum of
 integral for the increment and integral around the pole. Hence,
 the increment of phase is given by the residue at $\xi=\xi_+$,
\begin{equation}\label{inc-p02}
    \Delta_c\varphi_E=\pi.
\end{equation}
Remarkably, it is independent of $r_c$.

In \cite{K1,K2,K3} we have introduced a winding number,
determining a number of cycles per the full period of phase. By
\cite{K3} we recognize that the increment value of (\ref{inc-p02})
leads to the following conclusion: there is the only quantum level
of particle moving behind the horizon of BTZ black hole at $J=0$,
i.e. the level at $r_c\to r_+$ or $A\to -\infty$, since at
$r_c<r_+$ we expect the winding number equal to
$3\pi/(2\Delta_c\phi_E)=3/2$, which is not integer. The level is
the ground state of particle in the thermal ensemble. The
corresponding winding number is given by
\begin{equation}\label{wind}
    n=\frac{2\pi}{\Delta_c\varphi_E}=2.
\end{equation}
Thus, each particle can occupy two orbits at the ground level.

The increment of interval is determined by
\begin{equation}\label{inc-s0}
    \Delta_c s_0=2\int\limits_0^{r_c}{\rm
    d}r\,\frac{\ell}{\sqrt{r_c^2-r^2}}=\pi\,\ell,
\end{equation}
which is  also independent of $r_c$. Therefore, the action of
massive particle confined behind the horizon is equal to
\begin{equation}\label{act0}
    S_0=-mc\,n\,\Delta_c s_0=-mc\,2\pi\,\ell.
\end{equation}
Following \cite{K1,K2}, we deduce the thermodynamical function
\begin{equation}\label{G0}
    G=-\beta{\cal F}=-2\pi\,\ell\sum mc,
\end{equation}
where $\cal F$ is the Helmholtz free energy. Further, we denote
the sum of particle masses by
\begin{equation}\label{sig0}
    \sigma_0=\sum mc.
\end{equation}
Then, in the framework of thermodynamical relations, we fix the
total energy of the system by the black hole mass
\begin{equation}\label{mass0}
    M=-\frac{\partial G}{\partial
    \beta}=2\pi\ell\,\frac{\partial\sigma_0}{\partial\beta_0}.
\end{equation}
However, at $J=0$ we can easily use
$$
    \frac{\partial\beta_0}{\partial
    M}=-\frac{\pi\ell}{2M}\,\frac{1}{\sqrt{2GM}},
$$
so that
$$
    \frac{\partial\sigma_0}{\partial M}=-\frac{1}{4\sqrt{2GM}},
$$
that gives
\begin{equation}\label{sigma0}
    \sigma_0 =-\frac{1}{2}\sqrt{\frac{M}{2G}},
\end{equation}
where we have put the integration constant equal to zero, as
usually accepted. Then, the entropy can be evaluated by standard
formula of thermodynamics,
\begin{equation}\label{eS0}
    {\cal
    S}_0=\beta_0\,M+G=\pi\ell\sqrt{\frac{2M}{G}}=\frac{1}{4G}\,{\cal
    A}_0.
\end{equation}
The only unusual feature of such the derivation is the negative
sign of function $\sigma_0$ equal to the sum of particle masses.
This fact is closely connected with the notion of vacuum for the
black hole in the anti-de Sitter space (AdS). As was shown in
\cite{BTZ,Strominger}, the true vacuum is the AdS space
corresponding to the mass parameter equal to $M_{\rm
AdS}=-(8G)^{-1}$, i.e. the negative value of mass. This vacuum is
terated as a `bound' state. By analogy, the thermodynamical
ensemble has the negative free energy $\cal F$ at $J=0$,
$$
    {\cal F}_0 =-{M},
$$
that could be treated as the indication of binding the particles
in the field of BTZ black hole. If
\begin{equation}\label{<}
    {\cal F}_0\geqslant M_{\rm AdS}\quad \Rightarrow\quad
    M\leqslant\frac{1}{8G},\quad r_+\leqslant\ell,
\end{equation}
hence, the horizon has a radius less than the curvature radius of
AdS space.

The value of entropy is in agreement with calculations by
A.~Strominger in the framework of conformal theory
\cite{Strominger}. Indeed, the entropy counts the number of states
with the mass $M$, while the Virasoro algebra has the central
charge
\begin{equation}\label{c}
    c =\frac{3\ell}{2G},
\end{equation}
and generators
\begin{equation}\label{L0}
    L_0=\bar L_0 =\frac{2M}{\ell},
\end{equation}
so that the Cardy formula \cite{Cardy} gives
\begin{equation}\label{cardy0}
    {\cal S}_0 =4\pi\sqrt{\frac{c L_0}{6}}
    =\pi\ell\sqrt{\frac{2M}{G}}.
\end{equation}

The variation of $G$ can be written in the form
\begin{equation}\label{dG}
    {\rm d}G=\frac{\pi\ell}{2\sqrt{2GM}}\,{\rm d}M=\beta\,{\rm d}M,
\end{equation}
or
\begin{equation}\label{dG2}
    {\rm d}G=-2\pi\ell\,{\rm d}\sigma,
\end{equation}
which are consistent with each other. The Hawking radiation takes
place at ${\rm d}M<0$, i.e. at the decrease of black hole mass.
Eq. (\ref{dG}) means that the probability distribution versus the
energy emitted by the black hole $E=-{\rm d}M$, is given by
\begin{equation}\label{Gibbs}
    w\sim {\rm e}^{-\beta E},
\end{equation}
which is the Gibbs distribution, that reproduces the thermal
spectrum of black body up to `grey body' factors of re-scattering.
The radiation is balanced by the appropriate change of particle
masses at the ground level as dictated by (\ref{dG}) and
(\ref{dG2}),
$$
    {\rm d}\sigma=-\frac{\beta}{2\pi\ell}\,{\rm d}M.
$$
The consideration of Hawking radiation as a quasi-classical tunnel
effects was given in \cite{Wil}. Further details could be also
found in review on the thermodynamics of black hole horizons
\cite{Padma}.

\section{Two horizons: $\boldsymbol{J\neq 0}$}
At $J\neq 0$ we follow the consideration in second reference of
\cite{K2}. So, we put the angle dependent contribution in the BTZ
metric of  (\ref{BTZ}) equal to zero,
\begin{equation}\label{rotate}
    \dot \phi =\frac{4JG}{\rho^2},
\end{equation}
which corresponds to a definite relation between conserved total
energy $E$ and orbital momentum $\mu$. Then, we get the interval
\begin{equation}\label{BTZred}
    {\rm d}s^2=N^2{\rm d}^2t-\frac{1}{N^2}\,{\rm
    d}\rho^2
\end{equation}
in a strong analogy with the case of $J=0$. The difference is due
to the presence of two horizons and two return point for geodesics
completely confined behind the horizons.

The interval of trajectory takes the form
\begin{equation}\label{ds-traject}
    {\rm d}s^2=\frac{{\rm d}\rho^2}{{1}/{A}-N^2},
\end{equation}
where again $A=m^2/E^2$. For the confined geodesics $A<0$, and the
return points $\rho_{1,2}$ are positive solutions of equation
\begin{equation}\label{return-eq}
    \rho^4-\left(8GM\ell^2+\frac{\ell^2}{A}\right)\rho^2+16J^2G^2\ell^2=0.
\end{equation}
With $${\cal B}=1+\frac{1}{8GMA}$$ roots are
\begin{equation}\label{roots}
    \rho^2_{1,2}=4GM\ell^2\left[{\cal B}\pm\sqrt{
    {\cal B}^2-\frac{J^2}{M^2\ell^2}}\right],
\end{equation}
that indicates the existence of minimal value of negative $A$. The
roots are arranged, so that
$$
    \rho_-\leqslant \rho_1\leqslant\rho_2\leqslant\rho_+,
$$
of course.

The increment of interval is determined by
\begin{equation}\label{increm-s}
    \Delta_c s =2\int\limits_{\rho_1}^{\rho_2}{\rm
    d}\rho\,\frac{\ell\rho}{\sqrt{(\rho_2^2-\rho^2)(\rho^2-\rho_1^2)}}=\pi\ell.
\end{equation}
It is fixed and independent of return points.

The `tortoise' coordinate is given by
\begin{equation}\label{tortoise}
\begin{array}{rl}
    \rho_*= &\displaystyle\int{\rm
    d}\rho\,\frac{\ell^2\rho^2}{(\rho^2-\rho_+^2)(\rho^2-\rho_-^2)}=\\[7mm]
    &\displaystyle\hspace*{-2mm}
    \frac{\ell^2}{2(\rho_+^2-\rho_-^2)}\left[\rho_+
    \ln\frac{\rho-\rho_+}{\rho+\rho_+}-
    \rho_-\ln\frac{\rho-\rho_-}{\rho+\rho_-}\right],
\end{array}
\end{equation}
so that, after the introduction of appropriate Kruskal coordinates
and corresponding two maps with purely imaginary time \cite{K2},
we find two inverse temperatures of external and inner horizons,
correspondingly,
\begin{equation}\label{temper}
    \beta_\pm=2\pi\,\frac{\ell^2\rho_\pm}{\rho_+^2-\rho_-^2}.
\end{equation}
The consistency of two maps takes place, if
\begin{equation}\label{ratio-k}
    \frac{\beta_+}{\beta_-}=\frac{\rho_+}{\rho_-}={k},
\end{equation}
where $k$ is an integer or half-integer number \cite{K2'}.

The increment of phase defined by the purely imaginary time,
\begin{equation}\label{phaseE}
    \varphi_E=\frac{2\pi}{\beta_+}\,t_E,
\end{equation}
is given by the integral
\begin{equation}\label{increment-phiE}
    \Delta_c \varphi_E=
    \frac{2\pi}{\beta_+}\,2\int\limits_{\rho_1}^{\rho_2}{\rm
    d}\rho\, \frac{\sqrt{-1/A}}{N^2\sqrt{1/A-N^2}},
\end{equation}
which can be evaluated by the same way of previous section, i.e.
the contour integration after the transformation of variables
$\xi=\rho^2$. Then, it is reduced to contribution of two residues
by poles at $\rho^2=\rho_\pm^2$, where $N^2$ nullifies,
\begin{equation}\label{finincr}
    \Delta_c \varphi_E=\pi\,\frac{\rho_+-\rho_-}{\rho_+}.
\end{equation}
The winding number of ground level is equal to
\begin{equation}\label{winding}
    n_+=\frac{2\pi}{\Delta_c\varphi_E}=\frac{2k}{k-1}.
\end{equation}
It should be integer, that restricts values of $k$ by the
following set:
\begin{equation}\label{set-k}
    k=\left\{1,\frac{3}{2},2,3,\infty\right\}.
\end{equation}
$k=1$ represents the extremal black hole, while $k=\infty$
reproduces the black hole with $J=0$.

As we have already recognized in the previous section, there are
no excitations of ground state for the particle confined behind
the horizons of BTZ black hole.

Next, the action of single particle at the ground level is equal
to
\begin{equation}\label{action1}
    S=-mc\,n_+\,\Delta_c s=-mc\;
    2\pi\,\frac{\ell\rho_+}{\rho_+-\rho_-},
\end{equation}
or
$$
S=-mc\,\frac{\rho_++\rho_-}{\ell}\,\beta_+,
$$
which gives the partition function
\begin{equation}\label{partition}
    \ln Z=-\beta_+\,\sigma\,\frac{\rho_++\rho_-}{\ell},
\end{equation}
where $\sigma=\sum mc$ is the sum of particle masses ($c=1$).
Following the method of \cite{K2}, the total energy of system is
equal to the mass of black hole,
$$
    M=-\frac{\partial\ln Z}{\partial \beta_+},
$$
if
\begin{equation}\label{sumofmass}
    \sigma=\frac{\ell}{\rho_++\rho_-}\left(M-\frac{1}{4G\beta_+}\,{\cal
    A}_+\right),
\end{equation}
which reproduces the result of previous section for the sum of
masses at $\rho_+=r_+$ and $\rho_-=0$.

Further, in the standard thermodynamical way, the entropy
$$
    {\cal S}=\beta_+M+\ln Z
$$
is equal to
\begin{equation}\label{entr}
    {\cal S}=\frac{1}{4G}\,{\cal A}_+,
\end{equation}
in agreement with the Bekenstein--Hawking formula.

The Helmholtz energy
$$
{\cal F}= M-\frac{{\cal
A}_+}{4G\beta_+}=M\left(1-2\sqrt{1-\frac{J^2}{M^2\ell^2}}\right)
$$
is positive at
\begin{equation}\label{Helm}
    J^2\geqslant\frac{3}{4}\,M^2\ell^2\quad\Rightarrow\quad {\cal
    F}\geqslant 0.
\end{equation}

\section{Quantum spectrum}
The relation between the BTZ black hole mass $M$ and its orbital
momentum $J$ can be easily obtained from the condition of map
consistency (\ref{ratio-k}),
\begin{equation}\label{square-k}
    \frac{\rho_+^2}{\rho_-^2}=k^2\quad\Leftrightarrow\quad
    k^2 =\frac{1+\sqrt{1-\frac{J^2}{M^2\ell^2}}}
    {1-\sqrt{1-\frac{J^2}{M^2\ell^2}}},
\end{equation}
that gives\footnote{Here we put $J\geqslant 0$, for definiteness.}
\begin{equation}\label{solve}
    J=\frac{2k}{k^2+1}\,M\ell,
\end{equation}
which indicates the linear dependence of mass $M$ versus the
orbital momentum $J$. Then, the set of $k$ values determines 5
quasi-classical trajectories in the plane of $\{J,M\}$.

The set of winding numbers
\begin{equation}\label{n_+}
    n_+=\{\infty,6,4,3,2\},
\end{equation}
can be treated as numbers of massive spin-states: $n_+=2s+1$ (at
$J\neq 0$ and in non-extremal case).

Next, the `area' is also quantized as
\begin{equation}\label{a+}
    {\cal A}_+=2\pi\rho_+=4\pi\sqrt{G\ell J k}.
\end{equation}
It is spectacular that the `area' is not linear in integer or
half-integer $J$.

\section{Conclusion}
We have demonstrated that the approach of quasi-classical
quantization for the particles completely confined behind the
horizon \cite{K1,K2,K2',K3} can be successfully applied to the BTZ
black hole. The features of BTZ black hole are the followings: the
absence of excitations over the ground level, and the linear
dependence of mass versus the orbital momentum.

We could expect that the ground level of particles in the BTZ
black hole would be quite a coherent state, whose characteristics
at the external horizon is complete in a sense of holographic
principle \cite{Holograph}.

This work is partially supported by the grant of the president of
Russian Federation for scientific schools NSc-1303.2003.2, and the
Russian Foundation for Basic Research, grant 04-02-17530.


\end{document}